\documentstyle[prd,aps]{revtex} 
\begin{document}

\draft

\def\x{{\bf x}}
\def\y{{\bf y}}
\def\R{{\bf R}}

\newcount\equationno      \equationno=0
\newtoks\chapterno \xdef\chapterno{}
\def\eqn{\eqno\eqname}
\def\eqname#1{\global \advance \equationno by 1 \relax
\xdef#1{{\noexpand{\rm}(\chapterno\number\equationno)}}#1}

\def\la{\mathrel{\mathchoice {\vcenter{\offinterlineskip\halign{\hfil
$\displaystyle##$\hfil\cr<\cr\sim\cr}}}
{\vcenter{\offinterlineskip\halign{\hfil$\textstyle##$\hfil\cr<\cr\sim\cr}}}
{\vcenter{\offinterlineskip\halign{\hfil$\scriptstyle##$\hfil\cr<\cr\sim\cr}}}
{\vcenter{\offinterlineskip\halign{\hfil$\scriptscriptstyle##$\hfil\cr<\cr\sim\cr}}}}}

\def\ga{\mathrel{\mathchoice {\vcenter{\offinterlineskip\halign{\hfil
$\displaystyle##$\hfil\cr>\cr\sim\cr}}}
{\vcenter{\offinterlineskip\halign{\hfil$\textstyle##$\hfil\cr>\cr\sim\cr}}}
{\vcenter{\offinterlineskip\halign{\hfil$\scriptstyle##$\hfil\cr>\cr\sim\cr}}}
{\vcenter{\offinterlineskip\halign{\hfil$\scriptscriptstyle##$\hfil\cr>\cr\sim\cr}}}}}
\reversemarginpar

\title{\bf Event horizon: Magnifying glass for Planck length physics}
\author{T. Padmanabhan\thanks{paddy@iucaa.ernet.in}}
\address{IUCAA, Post Bag 4, Ganeshkhind, Pune 411 007, India.\\
IUCAA preprint 4/98 - January, 1998}
\maketitle

\begin{abstract}
 An attempt is made to describe the `thermodynamics'
of semiclassical spacetime without specifying
 the detailed `molecular structure' of the 
quantum spacetime, using the known properties of blackholes.
 I give detailed 
arguments, essentially based on the behaviour of quantum systems near
the event horizon, which suggest that  event horizon acts as a magnifying glass to probe Planck length 
physics even in those contexts in which the spacetime curvature
is arbitrarily low. 
The quantum state describing a blackhole,  in any microscopic description of spacetime, has to possess certain universal form of density of states which 
can be ascertained from general considerations.
Since a blackhole can be formed from the collapse of any physical system
with a low energy Hamiltonian $H$, it is  suggested
that when such a   system  collapses  to form 
a blackhole, 
it should be described by a modified Hamiltonian of the form $H^2_{\rm mod} =
A^2 \ln (1+ H^2/A^2)$ where $A^2 \propto E_P^2$.
 I also show that it is possible to construct several physical systems
which have the blackhole density of states and hence will be indistinguishable
from a blackhole as far as thermodynamic interactions are concerned. In
particular,  blackholes can be thought of as one-particle excitations of a class of {\it nonlocal} field theories
with the  thermodynamics of blackholes arising essentially
from the asymptotic form of the dispersion relation satisfied by 
these  excitations.  These field theoretic
models have correlation functions with a universal short distance
behaviour, which translates into the generic behaviour of semiclassical blackholes. Several
implications of this paradigm are discussed.   
\end{abstract}

\vskip 0.5in

\section {Introduction}

It is generally believed that the spacetime continuum will give way to
a more fundamental level of description at length scales smaller than
Planck length $L_P\equiv (G\hbar/c^3)^{1/2}$, corresponding to energy
scales larger than $E_P=L_P^{-1}$. Approaches to quantum gravity based on strings or Ashtekar
variables [1,2]  strengthens such a belief. If this is the
case, then spacetime continuum --- described by a solution to Einstein equations
--- is an approximate, coarse grained concept, similar to the continuum description
of a fluid or gas. Einstein equations have a status similar to that of 
equations of fluid mechanics and are of limited validity. 

Where exactly does the description based on Einstein's equations
fail? It seems reasonable
that the continuum description will fail whenever the length scale
associated with the spacetime curvature is comparable to that of Planck length. While this may be a {\it sufficient} criterion, it need not be {\it necessary}. It may be possible to see the effects of the underlying theory
in special circumstances, even if the curvature of the spacetime is arbitrarily
small. I will argue in this paper that physics near event horizons can give
 glimpses of the structure of the underlying theory, even if the
corresponding spacetime curvature is arbitrarily small, and show how this
information can be utilized.

The key reason which prompts me towards this point of view is the following.
If the spacetime has certain quantum mechanical micro-structure at 
Planck scales, and the continuum description based on Einstein's equations
is a coarse-grained one, then the {\it necessary} criterion for the breakdown (or otherwise)
of the continuum description should not be based on  the approximate theory.
An analogy might make this point of view clearer. In the study of a fluid 
system, one can obtain solutions to hydrodynamic equations describing the 
coarse-grained behaviour. In special circumstances, like in the case 
of shock waves, one can also obtain a {\it sufficient} condition for the 
breakdown of hydrodynamic description by studying these solutions. 
But if a very high energy beam of photons propagates through the fluid
probing length scales comparable to those of constituent particles, then
the smooth fluid description will necessarily breakdown around the region 
where the external influence interacts with the microscopic structure of 
the fluid. Similarly, if the  modes of an external field can probe
scales comparable to Planck length in some region of spacetime, then
we cannot describe physics around that region using Einstein's equations.
Normally, virtual modes of arbitrarily high energy of matter fields
do interact strongly with the quantum micro-structure of spacetime; but 
this is of no consequence unless such a virtual process can manifest 
as a real one in some way. This is exactly what happens in spacetimes
with an event horizon, like that of a blackhole. I shall elaborate on 
this theme in this paper.

This paper is organized as follows: Section II gives a detailed conceptual
foundation of this attempt and compares it with other approaches. Section III presents 
an argument indicating why the formulation of statistical mechanics of 
{\it any} system runs into trouble in spacetimes with horizon and suggests
a possible transmutation of the hamiltonian of a system, if the system
collapses to form a blackhole. The role of 
horizon in semiclassical gravity (ie., quantum fields interacting with 
classical gravity) is discussed next, in section IV. A simple derivation of 
Hawking evaporation is given, highlighting the role of infinite redshift surface. This analysis shows that event horizon is likely to play a vital role
in the quantum description of spacetime itself. This role is discussed in 
section V, where it is shown that all of blackhole thermodynamics can arise
from a particular form of density of states. In section VI, I discuss the
consequences of such a universal density of states for the blackhole and
present a simple field
theoretic model designed in such a way that the quanta of this field have this density of states. In this 
approach, blackhole represents the one-particle state of a given energy 
of this system. The last section summarizes the paper.
The key new results are in section VI and readers impatient with descriptions
of points of view may to go directly to that section, though it is not a recommended procedure.

\section {Prologue}

Since I want to attribute very special status to spacetimes with 
infinite redshift surfaces, I will begin by recalling certain well known 
features about them. To focus the ideas, I will discuss the case of a
Schwarzschild blackhole and use the concepts of infinite redshift surface
and event horizon interchangeably.

At the classical level, the horizon of  a blackhole (which is 
the simplest infinite redshift surface)  blocks out certain degrees of
freedom from the outside observers. Dynamically it allows one to describe
a blackhole by just three parameters irrespective of the initial configuration
of matter which collapsed to form the blackhole.
This idea has led Bekenstein [3] and later workers to suggest that ``blackholes
have entropy". Historically,  an entropy was attributed to the blackhole
before a temperature was  assigned, essentially due to the role played 
by the event horizon.

Since then many workers have tried [4] to understand the origin of blackhole
entropy and temperature but no clear consensus has emerged. The discussion 
usually centers around the the question of which degrees of freedom are
contributing to blackhole entropy.
Is the blackhole entropy a property of the matter fields which were 
outside the incipient blackhole ? Or, does it represent the 
degrees of freedom (both gravitational and matter)on the horizon or
inside, hidden by the horizon?

The  view that the entropy originates due to modes hidden by
the horizon is more in conformity with the idea that the entropy
arises from integrating over the degrees of freedom not relevant to the 
outside observer.  
On the other hand, the degrees of freedom of matter field outside the 
horizon are the ones which are of direct concern in any quantum field 
theory calculations carried out in Schwarzschild metric. The existence 
of event horizon or the specific nature of the metric enters only through 
boundary conditions imposed on the event horizon. Another way of 
stating this result is the following: The Feynman's Greens function for 
the quantum field contains all the information about the free fields 
which are
 outside the blackhole. Since it satisfies a local partial differential 
equation, it is independent of the detailed dynamics of the degrees of freedom
hidden inside the horizon. As far as the solution to that equation is concerned,
we  only need to impose a boundary condition on the event horizon.

A comparison of the above two arguments suggest that, in either case 
it is the 
{\it surface} of the blackhole, viz.the event horizon, which  will play a vital role. In the first line of 
thinking, this surface blocks out degrees of freedom and provides the 
justification for tracing them out. In the second approach, it is necessary
to impose boundary conditions on the surface in order to understand the 
dynamics of the fields in the outside. Again, in the first approach, it 
is very clear that this surface must be an event horizon to play this role; otherwise it cannot block out the degrees of freedom.
In the second approach, this is not so self-evident but arises
from the nature of the conical singularity at the horizon
and the structure of the corresponding Euclidian extension. There has also
been attempts in literature to attribute the entropy to the properties of
this surface itself [5]. 

In all the  approaches outlined above, quantum gravitational effects 
do not play any role and  one need not leave
the comforts of the spacetime continuum as a backdrop for studying physics.
(This is usually justified by arguing that the curvature at the event horizon
for a stellar mass black hole is quite small and quantum gravity should not
play a role; as I will argue in this paper, this is not sufficient justification.)
 A third point of view, which will be advocated here, is to treat the blackhole entropy as arising 
due to quantum structure of spacetime itself. In such an approach, one 
would like to think of classical spacetime as a very coarse level 
description (like that of a macroscopic gaseous system). The true 
Planckian level description of spacetime will involve certain degrees
of freedom which are excited  only when Planck energy processes
take place. Such a description, of course, should be universally applicable
irrespective of the nature of the  semiclassical or classical metric which eventually 
arises in a particular context. Since certain microscopic degrees 
of freedom of spacetime are traced out, one may think that
{\it any} spacetime will have a non zero entropy directly
related to the number of micro-states which are consistent with certain 
macroscopic parameters which identify the classical spacetime.
 The question then 
arises   as to what is special about a blackhole spacetime. The answer, as I will argue below,  has to do with  the existence of an infinite redshift 
surface. Classically, the infinite redshift of an event horizon acts 
as a great equalizer and arranges matters in such a way that very 
many collapsing configurations can all be described by the same kind
of metric. Semi-classically, the existence of infinite redshift immediately 
leads to  blackhole radiance and  thermal spectrum. At the next level,
quantum gravitationally, it is possible that the infinite redshift  
stretches out the subplankian degrees out of freedom and makes them 
`visible', to us at low energies. 

The above program has two parts: kinematical and dynamical. The
kinematical part attributes certain substructure 
to spacetime and assumes that any given macroscopic spacetime can be consistent
with very many different configurations of the underlying -- as yet unspecified -- substructure. This is, of course, similar to what is being done
in attempts to derive blackhole entropy from strings or Ashtekar variables [6,7]. But while these attempts are from ``bottom-up" and tries to arrive at
the entropy from a microscopic theory of quantum gravity, I want to proceed
in a ``top-down" fashion and try to arrive at some broad characterization
of the microscopic theory from the known classical and semiclassial 
properties of the blackholes.
In fact, the true spirit of the above approach will require 
 the final result to be independent of the detailed nature of the substructure
and  depend only on the existence of {\it some} substructure. For 
example, both strings and Ashtekar variables could lead to the same
kind of blackhole entropy. 

One clear (negative ?) aspect of this point of 
view, which seems to me inevitable, is that blackhole entropy is not 
going to help us in deciding the final version of quantum gravity.
Attempts to derive laws of quantum gravity using insights of blackhole 
entropy will be somewhat similar to attempts in deriving microscopic 
laws of physics from the laws of equilibrium thermodynamics. It should,
however, be remembered that original  hypothesis of Max Planck, $E=\hbar \omega$,  did arise from 
an attempt to understand the behaviour of radiation in thermodynamic
equilibrium. Something similar should be certainly possible. This could amount
to having an insight over the existence of certain internal degrees of 
freedom, their density of states  etc. Given the equilibrium thermodynamics of radiation 
field, one could have never obtained the full quantum theory of radiation, but
could only infer the existence of quanta. Similarly, given the equilibrium 
thermodynamics of blackhole, we will not be able to obtain the dynamical 
equations of quantum gravity but may be led to the 
inevitability of the existence of certain microscopic degrees of freedom which (eventually) leads 
to the continuum spacetime in the coarse grained point of view. 
On the positive side, such an approach has the  advantage that 
it is independent of the detailed dynamics of the microscopic theory about 
which we  have no consensus. We  should only require the minimal assumption 
that the microscopic structure of spacetime has certain degrees of freedom 
which are traced out in the macroscopic limit. The behaviour of a solid 
in terms of elastic constants or that of gas in terms of thermodynamic 
variables should not depend on our knowing the atomic physics; historically,
it did not. The question I want to address is similar: Can we develop
a working theory for `thermodynamics of spacetime' without knowing the
`molecular basis of spacetime'?

The dynamical aspect of the above program, which I will concentrate on in this paper, is closely related
to the task of  understanding
why spacetimes with infinite redshift are 
special. It is well known that frequencies of outgoing waves at late times
in blackhole evaporation 
correspond to super Planckian energies of the in-going modes near the horizon. One must then consider the possibility that these in-going modes  interact 
with the microscopic degrees of freedom of the spacetime located near the event horizon. This ties up  the two separate ideas: (i) There are internal degrees
of freedom in spacetime which are ultimately relevant for blackhole entropy. 
(ii) There is a dynamical mechanism for exciting these degrees of freedom in 
spacetimes with infinite redshift surface. The second feature is absolutely essential.
Existence of degrees of freedom which are unexcited in a given physical
context does not contribute to thermodynamics. For example, the degrees
of freedom of atomic nuclei or quarks  will not 
contribute to the specific heat of a solid at ordinary temperatures. This, 
in fact,  is the reason why the entropy of an ordinary spacetime
is effectively zero. In a normal spacetime, energy scales above Planck
energies are hardly excited and unexcited levels do not contribute 
to entropy. Infinite redshift surfaces provide a way of magnifying 
Planck level physics and make it visible. Dynamically, this arises from the 
fact that virtual modes of Planckian energy are converted into 
real modes of subplankian energies in such spacetimes. I stress the fact
that having spacetime micro-structure is only one aspect of any entropy 
calculation. One should have reasonable grounds to believe that dynamical 
process which excite these micro levels are also present in the spacetime. 
This happens naturally when an infinite redshift surface is available.

It is also easy to see how certain well known features will arise naturally
in this picture. Since one cannot
operationally measure
 length and
time scales below Planck values, any complete theory necessarily
has to accommodate this feature in a fundamental manner. In the naivest level, this  is equivalent to assuming that the 
number of independent micro-cells which are present in a region of spacetime of volume 
$V$ will be about $N\approx V/L_P^3$. 
All these states will {\it not}, however, 
contribute to entropy in the semiclassical limit. In order to make these  
states contribute it is necessary to have a physical process which excites 
them. Though virtual processes of Planckian energies and above will be constantly
exciting and de exciting them,  to get {\it observable} entropy, we need a situation in which the
{\it virtual} excitation of super Planckian energies are converted
 to {\it real} 
excitations of sub Planckian energies. This is possible in a spacetime with
an infinite redshift surface, by the in-going modes of super Planckian energies.
But it occurs in a  thin shell of thickness $L_P$ around the 
event horizon, i.e., in a region of volume  of the size $V \cong A L_P$
where $A$ is the area of the event horizon. 
 Clearly the entropy will now
turn out to be proportional to $N\cong (V/L_P^3) = (A/L_P^2)$, that is, to the area of the infinite redshift surface.

\section{ Phase volume and entropy in spacetimes with horizons}
 
Since the  event horizon  seems to play an important role in the study of  statistical concept like entropy, I will begin by discussing a thought
experiment related to the statistical description of any system in a 
region surrounding an incipient blackhole. The statistical mechanics of 
such a system will require computation of the phase volume available for 
the system when its energy is $E$. I will show that certain difficulties 
arise in formulating the statistical mechanics of the system even at the 
classical level when an event horizon forms in the spacetime.

 Consider a system
of $N$ relativistic point particles located in a 
static spacetime with line element
\begin{equation} 
ds^2  = g_{00}({\bf x})dt^2 - \gamma_{\alpha \beta}({\bf x}) dx^\alpha dx^\beta \qquad (\alpha, \beta = 1,2,3)\label{eqn:qmetric}
\end{equation}
Any particle with four-momentum $p^a$ possess the conserved energy (scalar) $E = \xi_ap^a$ in such a spacetime where $\xi^a = (1, {\bf 0})$ is 
the time-like killing vector field.  
 Standard statistical mechanics of this 
system is based on the density of states $g_{N}(E)$ which, in turn,
 can be built out of the density of state
$g(E)\equiv g_{N=1}(E)$ for individual particles;
\begin{equation}
 g_N(E) = \int \Pi^N_i \, g(E_i) \, dE_i \, \delta(E - \sum E_i)
\end{equation}
The volume of phase space available for  a single particle with 
energy {\it less than} $E$, is given by
\begin{equation}
\Gamma (E) = \int dx^\alpha dp_\alpha\ \Theta(E - \xi^ip_i) = {4\pi \over 3}  \int \sqrt{\gamma} \, d^3x^\alpha \, (E^2 / g_{00} - m^2 )^{3/2}
\end{equation}
where $\Theta $ is the Heavside theta function. 
The first form of the expression shows that $\Gamma (E)$ is generally 
covariant; the second expression is obtained by integrating over 
the momentum variables. 
This result is identical to the one which would have been used by
any locally inertial observer at an event ${\cal P}(t, x^\alpha)$ who 
attributes a local value of energy $E_{\rm loc} =u^i p_i = E(g_{00})^{-1/2}$
to the particle. (For more details, see [8].) The density of states
is given by $g(E) = (d\Gamma(E)/dE)$ and the entropy of the system
is $S(E) = \ln g(E)$. 
All of the standard statistical mechanics of the system in an external 
gravitational field described by the metric in (\ref{eqn:qmetric}) will follow
from the expression for the space volume.

Consider now the spacetime of a spherical star of 
mass $M$ and radius $R > 2M$. The phase volume available for a 
particle located outside the star  is given by
\begin{equation}
\Gamma (E) = {16\pi^2\over 3} \int_R^{R_{\rm max}} {r^2 dr\over (1 - 2M/r)^2}  \left[ E^2  - m^2\left(1-{2M\over r}\right) \right]^{3/2}
\end{equation}
where $R_{\rm max}=2M(1-E^2/m^2)^{-1}$ is the maximum radius allowed for energy $E<m$. For $E>m$, we shall assume that the system is confined inside
a large volume $V$ as is usual in statistical mechanics. This 
expression, as well as the statistical mechanics developed by using it,
remains well defined as long as $R >2M$ but diverges when $R\to 2M$. Let us suppose that
 the star now starts to collapse eventually forming a blackhole.
At late times $t\ga 2M$ the radius of the star will follow the trajectory (see
e.g., ref.[9]):
\begin{equation}
R(t) \cong 2M + 2M\,  e^{-t/2M}
\end{equation}
making the phase volume $\Gamma(E)$ and  density of states $g(E)$ diverge exponentially in time; for example,
\begin{equation}
 g(E) = {d\Gamma\over dE} \cong 16 \pi^2 E^2 (2M)^3 \exp \left( {t\over 2M}\right)
\end{equation}
for $t\ga 2M$. 
In other words, the entropy of {\it any} system of particles located
outside a collapsing star diverges as the star collapses to form
a blackhole due to the availability of infinite amount of phase
space. 

Similar disaster occurs even for a field. The wave equation
describing a scalar field can be written in the form
\begin{equation}
\left( {\partial^2 \phi\over \partial t^2} - {1\over 4 M^2}{\partial^2 \phi \over \partial x^2} \right) + e^x \left( - {1\over 4 M^2} \nabla_{(2)}^2 \phi + m^2 \phi \right) = 0 
\end{equation}
near the horizon if we use the spatial coordinate $x$ defined through
$r-2M =2M \exp(x)$. (Here $\nabla_{(2)}^2$ is the Laplacian on the $r = $ constant surface). As $x\to -\infty$, the field becomes free 
and solutions are simple plane waves propagating
all the way to $x = -\infty$. The existence of such a continuum
of wave modes will lead to infinite phase volume for the system.
More formally, the number of modes  $N(E)$ for a scalar field $\phi$ with vanishing
boundary conditions at two radii $r=R$ and $r=L$ is given by
\begin{eqnarray}  N(E) & &\displaystyle{ \cong {1\over \pi} \int^L_R {dr\over \left( 1 - 2M/r\right)} \int dl\, (2l+1) \left[ E^2 - \left( 1 - {2M\over r}\right) \left( m^2 + {l(l+1)\over r^2} \right) \right]^{1/2}} \nonumber \\
& &\displaystyle{ = {2\over 3\pi} \int^L_R {r^2 dr\over \left( 1 - 2M/r\right)^2}\left[ E^2 - \left( 1 - {2M\over r}\right)  m^2\right]^{3/2} ={\Gamma(E)\over(2\pi)^3}}\label{eqn:qnofe}
\end{eqnarray}
in the WKB limit [10,11]. This expression diverges as $R\to 2M$ showing that
a scalar field propagating  in a blackhole spacetime has infinite phase volume and entropy.
The divergences described above occur around any infinite
redshift surface and is a geometric (covariant) phenomenon.

This result, which shows that even conventional statistical mechanics does not
exist in a spacetime with infinite redshift surface, need to be taken seriously.
Physically, blackholes are fairly strange kind of entities when viewed
from the point of view of statistical mechanics; statistical physics is 
based on {\it mutual} interactions of constituents leading to energy exchange
where any subsystem can influence {\it and} be influenced by the rest of the 
system. Regions of spacetime covered by infinite redshift surfaces
break this symmetry, since the material  has fallen inside the horizon  cannot causally influence
the outside.

The key {\it mathematical} reason for all the above divergences is the behaviour of locally
defined energy $E_{\rm loc}=E/(g_{00})^{1/2}$ at the surface of the star $r=R$ as $R\to 2M$; in this limit,
$E_{\rm loc}\to \infty$ and the available phase volume becomes infinite essentially
due to the behaviour of the system near the horizon. This feature suggests that
the  solution to this problem lies in physics at arbitrarily high energies.
If true, there is hope for turning around this argument in order to 
catch a glimpse of Planck scale physics by demanding that the breakdown 
of statistical mechanics should not occur.

It is very likely that the 
 classical divergence of phase volume signals the necessity of a 
new physical principle in dealing with blackholes.
Recall that in classical relativity a Schwarzschild blackhole can be formed out of 
any spherically symmetric  system provided sufficient amount of energy is confined to 
a small enough radius. Consider an arbitrary physical system with a 
Hamiltonian $H$ which is collapsing in a spherically symmetric manner 
to form a blackhole. If the initial 
energy of the configuration is $E$, which will be conserved 
during a spherically symmetric collapse, the resulting blackhole
will have  a mass $M=E$. Initially, the system has access to a phase volume
$\Gamma(E)$ which depends only on the form of the Hamiltonian $H$ and the
energy $E$. At sufficiently late times, the radius of the system
approaches $R \to 2M$ with the energy remaining the same. If we assume 
that the Hamiltonian describing a system is the same irrespective of whether 
the system becomes a blackhole or not, it follows that the phase volume
available to the system diverges as $R\to 2M$ but the formal dependence
of $\Gamma(E)$ on $E$ remains the  same. (In the example studied above, 
$\Gamma(E) \propto E^2 f(t)$ with $f(t)\to \infty$ for $R\to 2M$; the $E$ dependence is a power law.). We shall, however, see in section 5 that blackholes should be described by a universal form of 
phase volume $\Gamma_{\rm BH}(E) =\exp(4\pi E^2/E_P^2)$ quantum mechanically.
[This phase volume is very large for macroscopic blackholes with $E\gg E_P$
but is finite due to quantum mechanical considerations. We see the historically familiar phenomena of a classical divergence being 
regularized by quantum mechanics, with the replacement of infinity by a large number.]
But the important point to note is that the energy dependence of the 
phase volume has to change completely when the system collapses to form a black hole with $R \to 2E$. 
If the energy is conserved, then such a change can occur only if 
the Hamiltonian of the system gets transformed (by, as yet unknown, quantum
gravitational process) when the system forms a blackhole. If I postulate
that the formation of a blackhole, from a 
conventional physical system with hamiltonian $H_{\rm conv}$, changes the
hamiltonian to the form $H_{\rm mod}$, with
\begin{equation}
 H_{\rm mod}^2=A^2\ln\left(1+{H^2_{\rm conv}\over A^2}\right); \qquad A\propto E_P
\end{equation}
then, the formation of the blackhole will lead to a universal form of 
phase volume $\Gamma_{\rm BH}(E)$ at the conserved value of the
energy $H_{\rm conv} =E$. The arguments given later in section 6
suggest that such a transformation has certain inevitability
and leads to interesting consequences.

\section {Event horizon in the semiclassical limit} 

 As shown above, the existence of event horizon leads to difficulties in formulating statistical mechanics for {\it any} system even classically. 
At the next level, when one studies quantum
filed theory in the blackhole spacetime, the event horizon assumes a greater role: If the event horizon is formed due to collapse of matter, then
it radiates like a black body of temperature $T=(8\pi M)^{-1}$ at late times $t \gg 2M$.
It is precisely the existence of an infinite redshift surface which leads to the
characteristic Planckian form of the spectrum and distinguishes the blackhole
of mass $M$ from, say, a neutron star of mass $M$.  While this feature is apparent
in the primary derivation of blackhole radiation by Hawking, it is worthwhile
to establish the connection between infinite redshift surface and the Planck
radiation in a simple and direct manner which I shall now do.

Consider a radial null geodesic in the Schwarzschild spacetime which propagates
from $r=2M+\epsilon$ at $t=t_{in}$ to the event ${\cal P}(t,r)$ where
$\epsilon\ll 2M$ and $r\gg 2M$. The trajectory can be determined from the
Hamilton-Jacobi equation for the action ${\cal A}(r,t)\equiv -Et +S(r)$, written
in the form
\begin{equation}
g^{ik} \partial_i {\cal A}\,  \partial_k{\cal A} = {E^2\over \left( 1 - {2M\over r}\right)} - \left( 1 - {2M\over r}\right)\left({dS\over dr}\right)^2 =0
\end{equation}
The trajectory, determined by the condition $(\partial {\cal A}/\partial E)=$
constant,  with the required initial conditions is
\begin{equation}
 r \cong t - t_{\rm in} + 2M \ln \left( {\epsilon\over 2M}\right) \qquad (\epsilon\ll 2M,\quad r\gg 2M)\label{eqn:qeight}
\end{equation}
The frequency of a wave will be redshifted as it propagates on this trajectory.
The frequency $\omega$ at $r$ will be related to the frequency $\omega_{in}$
at $r=2M+\epsilon$ by 
\begin{equation}
\omega \cong \omega_{\rm in} \left[ g_{00} (r=2M+\epsilon)\right]^{1/2} \cong \omega_{\rm in} \left({\epsilon\over 2M}\right)^{1/2} = \omega_{\rm in} \exp\left( -{t - t_{\rm in} -r \over 4M} \right) \label{eqn:qinsom}
\end{equation}
where we have used (\ref{eqn:qeight}).  If the wave packet, $\Phi(r,t)\simeq \exp(i\theta(t,r))$, centered on this null ray has a phase $\theta(t,r)$, then
the instantaneous frequency is related to the phase by $(\partial \theta/\partial t)=\omega$.
Integrating (\ref{eqn:qinsom}) with respect to $t$, we get the relevant wave mode to be
\begin{equation}
\Phi(t,r) \propto \exp i\int \omega \, dt \propto \exp \left[ - 4 M \omega i \ \exp\left( - {t-t_{\rm in} -r \over 4M} \right)\right] 
\end{equation}
(This form of the wave can also be obtained by directly integrating the 
wave equation in Schwarzschild geometry with appropriate boundary conditions; 
the above derivation, however, is simpler.)
An observer using the time coordinate $t$ will Fourier decompose these
modes with respect to the frequency $\nu$ defined using $t$:
\begin{equation}
\Phi(t,r) = \int^\infty_{-\infty} {d\nu\over 2\pi} f(\nu) e^{-i\nu t}
\end{equation}
where
\begin{equation}
 f(\nu) = \int_{-\infty}^\infty dt \, \Phi(t,r) e^{-i\nu t} \propto \int_0^\infty dx\, x^{- 4 M \nu i - 1} \exp(-4M \omega_{\rm in} ix) 
\end{equation}
The integral can be evaluated by rotating the contour to the imaginary axis.
The corresponding power spectrum is
\begin{equation}
|f(\nu)|^2 \propto \left( \exp (8\pi M\nu) - 1 \right)^{-1} \label{eqn:qpsp}
\end{equation}
which is Planckian at temperature $T=(8\pi M)^{-1}$. 

The simple derivation given above strips the process of Hawking evaporation to its bare bones
and establishes the following: (i) The key input which leads to the Planckian
spectrum is the exponential redshift  given by equation (\ref{eqn:qinsom})  of modes which scatter off the blackhole and travel to infinity at late times. This requires
an infinite redshift surface and --- in fact --- the derivation can be easily
generalized to other cases which have infinite redshift surfaces (Rindler,
de Sitter ...). (ii) The analysis up to equation (\ref{eqn:qpsp}) is entirely classical
and no $\hbar$ appears anywhere. In normal units, $8\pi M\nu$ becomes $8\pi GM\nu/c^3$. It is only our desire to introduce an energy $E=\hbar\nu$ which
makes this factor being written as $8\pi GM\nu/c^3=(8\pi GM/\hbar c^3)\hbar\nu
=E/k_BT$. The mathematics of Hawking evaporation is purely classical and lies
in the Fourier transform of an exponentially redshifted wave mode (for a discussion of the classical versus quantum features see ref.[12]).

It is clear from the above derivation that  the existence of infinite redshift surface  allows 
modes with frequencies $\omega \gg E_P$ near the horizon to appear as sub Planckian radiation at future null infinity. Any
non trivial part of the Planck spectrum arising out of Hawking 
evaporation will correspond to   a mode which originated with virtual energies much higher than
 Planck energy near the horizon. For example, frequencies near the peak
of the Planck spectrum, $\omega\approx (8\pi M)^{-1}$ will arise from
incoming modes with frequencies $\omega>E_P$ at all retarded times $u> t_{Q}\equiv 4M\ln (8\pi M/E_P)$. For a stellar mass blackhole $(t_Q/2M)\approx 10^2$. Since the Planck spectrum gets established and the transients die down only for $t\ga 2M$,
most of the Hawking radiation originates from modes with $\omega\gg E_P$ near 
the horizon.
These modes also approach the event horizon arbitrarily closely; for $u>t_Q$
the modes have originated from $2M<r<2M+L_P$.
 Clearly, the existence of a horizon
allows one, through the process of blackhole evaporation,
to probe the physics at sub Planckian length scales near the horizon. 

We have now seen two examples of how arbitrarily high energies near 
the event horizon plays a vital role. In the case of phase volume this 
creates difficulties in formulating a viable statistical mechanics
and makes the entropy of any system formally infinite. In the case of 
Hawking evaporation, these modes are primarily responsible for most 
of the Planck spectrum seen at late retarded times. These two features 
 suggest that 
 any attempt to obtain a correct description of blackhole
evaporation will be incomplete without addressing the arbitrarily 
high frequency virtual excitations near the horizon. Clearly, the relevant 
issue is not whether the gravitational field described by Einstein's
equation has arbitrarily high curvature; the curvature at the event 
horizon of a stellar mass blackhole can be quite low. The relevant 
question, it appears to me, is whether processes in any local region involves
energy higher than Planck energies. If they do, the description in 
terms of continuum spacetime breaks down and we have to worry about
Planck scale physics.  
 
 Let us ask what this paradigm implies for the divergence of the phase
volume derived earlier. To begin with, it is clear that the problem is ``local" and confined to a
region around the event horizon. Since the energy of the modes are higher
than Planck energy as one approaches the horizon, it is  likely that
the excitation of microscopic degrees of freedom of spacetime (whatever they
may be) will occur in this region.  Let us suppose that
 the  modes of the scalar field excites the microscopic
degrees of freedom of spacetime in a shell-like region between $r_1=2M+h$ and $r_2=2M+H$.
The {\it proper} distance $l$ of these surfaces from $r=2M$ will be taken to be
some definite multiple of Planck length with $l_1\equiv (L_P/c_1)$ and $l_2\equiv c_2 L_P$ where $c_1$ and $c_2$ are,
at present, unknown numerical constants. 
(We expect both $c_1$ and $c_2$ to be larger than unity so that $l_1 < L_P$ and $l_2 > L_P$.)
It is easy to relate the proper
and coordinate distances and obtain, for $L_P\ll 2M$,
\begin{equation}
h \cong (L_P^2/8M) \ c_1^{-2}; \qquad H \cong (L_P^2/8M)\  c_2^2\label{eqn:qprol}
\end{equation}
The modes of the scalar field in this region interacts with the microscopic
degrees of freedom and I will now assume that this interaction drives the system to a state of maximum probability. In other words
local thermodynamic equilibrium exists between the degrees of freedom 
of the spacetime and the modes of the scalar field leading to some
local temperature $T/(g_{00})^{1/2}$ (The corresponding value of temperature
at infinity is $T$). The entropy of the scalar modes confined to this region
at some given temperature  is straightforward to compute using the standard relations:
\begin{equation} S = \beta \left[ {\partial \over \partial \beta} - 1 \right] F; \qquad
F = - \int_0^\infty dE {N(E)\over e^{\beta E} - 1 },
\end{equation}
and the expression for
 $N(E)$ in equation
(\ref{eqn:qnofe}) (see  ref [10,11]). We get
\begin{equation}
 S = {32\pi^3\over 45} (2MT)^3 \left({2M\over L_P}\right)^2 \left\{ c_1^2 - {1\over c_2^2} \right\} \label{eqn:qsix}
\end{equation}
When the modes of the scalar field propagate to infinity, they are
redshifted to sub Planckian energies and appears with the spectrum in (\ref{eqn:qpsp}), corresponding to the temperature
at infinity to be $T=1/8\pi M$. Using this in (\ref{eqn:qsix}) we get
\begin{equation}
S = 4\pi\,  {M^2\over L_P^2} \left[ {1\over 90 \pi} \left( c_1^2 - {1\over c_2^2}\right)\right]\label{eqn:qseven}
\end{equation}
There are several features to note in this expression. (i) To begin with
it shows that $c_2$ is relatively unimportant and we could as well take
$c_2\to \infty$; this is to be expected since most of the divergent
contribution arises from the lower limit of the integration. Hereafter, we shall set $c_2^{-1} =0$. (ii) Second, the
result diverges when $c_1 \to \infty$; this is just the old result of the divergence
of phase volume for $R\to 2M$. Arguments I gave above, however,  suggests that we 
are not allowed to take this limit since it makes the locally defined energy,  $E_{\rm loc}$, of 
the mode to exceed the Planck energy by an arbitrary amount.
I will discuss  the actual value of $c_1$ later on.  
(iii) Third, expression (\ref{eqn:qseven}) gives an entropy which is proportional to
the area of the event horizon. This is non trivial and two key mathematical 
features of the spacetime geometry have conspired to produce this 
result: (a) The metric coefficient $g_{00}$ is the reciprocal of
$g_{11}$ and $g_{00}$ goes as ${\cal R}(r-r_H)$ near  the event horizon
 $r_H = 2M$.  It is easy to show that any such 
spacetime will lead to a thermal state with temperature $T = ({\cal R}/4\pi)$.
(This is proved, for example, in ref.[13] and is directly verifiable
from the  periodicity in Euclidean time.) (b) The coordinate distance
and the proper distance to the event horizon scales as in equation (\ref{eqn:qprol}).
This, in turn, is a consequence of the metric having a simple zero 
at the horizon.  The proportionality between the 
entropy and the area will fail except when the above two mathematical criteria are 
met. In other words, the result holds only around an infinite redshift 
surface and not at any arbitrary radius.

The physical reason for this 
proportionality to area is simple  and more important. In the approach 
I have outlined,  any volume $L^3$ in space
is assumed to be made of $N=(L/L_P)^3$ Planck sized cells.
The microscopic degrees of freedom in these cells can be 
excited {\it only} when (i) sufficiently high virtual energies are available
and (ii) some mechanism exists for these modes to convert themselves as low 
energy propagating waves. The first condition is met  only in a small region 
of thickness of the order of $L_P$ around an infinite redshift
surface. Hence the number of microscopic cells which can contribute
to the entropy is reduced from the total number of cells inside the 
blackhole, $N_{\rm in} = (2M/L_P)^3$ to those residing in a thin shell of radius
$r=2M$ and thickness $L_P$; the latter is about $N_{\rm shell} \cong (2M)^2(L_P)/L_P^3 =(2M/L_P)^2$.

The {\it actual} value of the entropy contributed by the scalar field
cannot be computed without knowing the value of $c_1$.
If we take $c_1 = (90\pi)^{1/2}$, then the entropy turns out to  be one-fourth
of the area of the event horizon which is the Bekenstein-Hawking result.
In the conventional discussions of blackhole entropy, a sharp 
distinction is made as to whether the entropy is contributed
by (a) non gravitational degrees of freedom external to blackhole or 
(b) those internal to the blackhole or whether (c) the entropy is a basic 
property of the gravitational field of the blackhole itself. In the 
 approach advocated here, this question becomes somewhat irrelevant. Near the 
event horizon,  high energy excitations of the scalar field interact 
strongly with some unspecified degrees of freedom of quantum spacetime. The 
entropy contained in this region of strong coupling, cannot be properly 
separated as entropy of either the scalar field or that of the blackhole.
The blackhole constantly scatters the incoming modes to outgoing modes
which should  be thought of as excitations followed by a radiative
decay of the microscopic degrees of freedom of spacetime around the 
event horizon.  

In the thermodynamic interaction between any two systems, temperatures
will be equalized when the system is driven to the state of maximal
space volume. I will argue in the next section that  the density of state of microscopic spacetime degrees
of freedom in the case of a blackhole has the form $g(E) = \exp[4\pi(E/E_P)^2]$.
Given this,
we are assured that any other field which come into interaction
with these degrees of freedom near the horizon will behave as 
though it has a temperature $T = (8\pi M)^{-1}$. Actually this is
{\it all} we know from the semiclassical analysis of quantum fields in curved
spacetime. The total entropy of the system made of microscopic spacetime degrees
of freedom as well as  the matter fields is not an operationally well 
defined concept. If some fundamental theory of quantum spacetime provides
the density of states $g(E) = \exp[4\pi(E/E_P)^2]$, rest of blackhole
thermodynamics will follow from such a result. From this point of view, 
the actual value of $c_1$ is quite irrelevant and we need {\it not} arrange it
so as to get any particular value for the  entropy.

In this approach, the density of states of the microscopic degrees of 
freedom is a more fundamental construct than the so called entropy
of blackhole. The role of a blackhole is limited to providing a simple
context in which these microscopic degrees of freedom can be excited and 
de-excited due to the existence of an infinite redshift surface. In this
sense blackholes act as a magnifying glass which allows us to see
the microscopic quantum spacetime.
[The mathematics of this 
analysis was first given in the `brick-wall model' of t'hooft [10] and 
 was generalized in ref. [11]; the interpretation given here, however, 
is quite different.]

\section{ Event horizon and micro-structure of spacetime}

Having shown that the interaction between quantum micro-structure of spacetime and the local high energy modes of a scalar field can occur near an 
event horizon, I now take up the question of the nature of these
microscopic degrees of freedom. As emphasized several times before,
there is no way one can obtain the detailed theory of quantum gravity
just from the results derived so far. The hope is to obtain
some general characteristics of blackhole spacetimes which is 
enough to provide a description of the results obtained 
above. I will first re-derive a result obtained earlier by t'hooft
for the density of states of a blackhole and  will then show how
 one can construct an effective field theory from which blackhole
spacetime will emerge as an ``one-particle'' excitation.  

There are
several reasons to believe that event horizon will continue to play a special role in the next level of
description --- viz. in a microscopic theory of spacetime.  Any classical
(asymptotically flat) spacetime with energy $M\gg E_P$ has to arise
as some nonperturbative quantum condensate of the true microscopic degrees of freedom (strings, membranes, spin networks....). The classical solutions
for a spherically symmetric neutron star of mass $M$ and for a blackhole of mass $M$ belong to this class. But there is a fundamental difference between these two solutions: A neutron star is stable against Hawking evaporation while a blackhole is not. The quantum states representing
these two objects must reflect this feature in some manner. Naively, one
would expect the energy of the state to pick up an imaginary part characterizing
the decay. If $|M, {\rm star}> $ and $|M, {\rm bh}>$ are the two quantum states, then
at some suitable limit, the time evolution of these states will be
\begin{equation}
|M,t,{\rm star}>=\exp (-iMt)|M,0,{\rm star}>;\quad |M,t,{\rm bh}>=\exp\left( -i[M -iQ(M)]t\right] |M,0,{\rm bh}>
\end{equation}
where $Q(M)\propto (E_P/M)^2 M^{-1}$ is the adiabatic decay rate of the energy
due to Hawking evaporation. How can two states, which are parametrized classically by the same quantity $M$ have very different quantum descriptions ?
This is possible {\it only if} the microscopic  theory  takes cognizance of
the existence of the infinite redshift surface in one of the solutions,
 since these two
classical solutions differ only in that aspect. Hence the microscopic theory
must have a mechanism to take this feature into account
 even though
the issue is not directly related to the existence of high curvature.

Consider one such quantum state, describing a blackhole of mass $M$ (say).
Whatever may be
the microscopic degrees of freedom, we can try to characterize their configuration in such a state 
by their average energy $\bar E$ and density of states $g(\bar E)$. Given $g(\bar E)$, we can define the entropy $S(\bar E) \equiv \ln g(\bar E)$ and inverse  temperature $\beta (\bar E) = (\partial S / \partial \bar E)$. Note that these are well defined, formal constructs defined using microcanonical
ensemble, without using any external systems or heat bath. We are interested  in the form of $\beta(\bar E)$.
Since $\bar E$ has to be determined by the spacetime geometry in 
the classical limit,  it 
appears quite reasonable to take $\bar E = M$. (If the Hawking 
evaporation is thought of as the decay of an unstable quantum 
state, then it is {\it necessary} that $\bar E =M$.)  
The temperature is related to $M$ by $T=(E_P^2/8\pi M)$, thereby
leading to the result  $\bar E = (\beta/8\pi)$. 

I want to 
stress that a physical system with such a relation between mean 
energy and temperature  must possess very unconventional 
features. Conventional statistical mechanics invariably leads to
a mean energy which is an increasing function of temperature. This
is trivially true for any system possessing canonical ensemble description since
the specific heat in canonical ensemble must be positive definite. But even
ordinary self gravitating systems have negative specific heats and do not
allow a description in terms of canonical ensemble. In the micro-canonical
ensemble, it is possible to have negative specific heats with mean energy
decreasing with temperature. One simple consequence of this fact is that when two blackholes combine to
form a larger one, the total energy (and density of states) increase
but the temperature  decreases. This suggests that the degrees of 
freedom of spacetime from which blackhole states are built  should be described  by micro-canonical
ensemble and its partition function may exist only in a formal sense.

Let us next consider another feature of the quantum state describing a 
blackhole, namely, the density of states. Since the macroscopic description
uses only the mean energy and does not specify the detailed configuration 
of the microscopic quantum degrees of freedom of the spacetime, it stands
to reason that the quantum states will be highly degenerate and should 
be describable in terms of some density of state function $g(\bar E)$.
The fact that entropy $S$ is proportional to the area (${\cal A} \propto M^2$) alone suggests that 
the density of states for a blackhole of energy $E=M$ scales as
$g(E) = \exp(S(E)) = \exp(\alpha_1 E^2)$ where $\alpha_1 $ is some constant. This is very
different from the scaling which arises in statistical mechanics of 
normal systems. For a system made of massless quanta at temperature $T$, we have
$S \propto T^3, E\propto T^4$ giving $S\propto E^{3/4}$; hence 
the density of states grows  as $\exp[\alpha_2(LE)^{3/4}] $
where $L$ is the linear dimension of the system and $\alpha_2$ is 
some constant. If such a system (with energy $E$ and confined in a spatial 
region of size $L$) should not become a blackhole, then we must have
$E<(L/L_P^2)$. Using this bound we see that
the density of states for such a system is bounded by a form
$\exp[\alpha_3(L/L_P)^{3/2}]$. This increases more slowly than the form $\exp[({\rm constant})\ L^2]$.

Such a rapid growth of the phase volume, $g\propto \exp(\alpha_1 E^2)$,  immediately implies that one cannot
obtain the partition function as a standard Laplace transform of the 
density of states. This, however, should not come as a surprise since
the existence of a well defined partition will imply the existence 
of a description in terms of  canonical ensemble and consequently 
positive specific heat. Since the blackhole state has negative 
specific heat, something has to give way and what breaks down 
is the conventional relation between $g(E)$ and $Z(\beta)$. We 
can, however, obtain a partition function by a different procedure as follows.
We consider the density of states
$g(E)$ of a blackhole to be a function of a complex variable $z$
and study the behaviour of $g(z)$ along the imaginary axis $z=iy$.
If $g(z)\propto \exp[4\pi (z/E_P)^2]$, it remains bounded along the 
purely imaginary axis with $g(y) \propto \exp[-4\pi(y/L_p)^2]$.
Defining the partition function $Z(\beta)$ as  Laplace transform along the 
imaginary axis (which becomes a Fourier transform), we get
\begin{equation}
Z(\beta) = \int_{-\infty}^\infty dy\, \rho(y) e^{-i\beta y} \propto \exp\left( - {\beta^2 E_P^2\over 16\pi}\right) 
\end{equation}
Such a partition function will correctly reproduce the relation between
mean energy and temperature
\begin{equation}
 M = \bar E = - {\partial \ln Z(\beta)\over \partial \beta} = {E_P^2 \over 8\pi} \beta = {E_P^2\over 8\pi T}
\end{equation}
which justifies the formal manipulations a posteriori.

The above analysis shows that the key feature characterizing the microscopic
quantum degrees of freedom of spacetime with event horizon  is the peculiar density of states
of the form $g(E) = \exp[4\pi(E/E_P)^2]$ [This form for $g(E)$ was earlier
derived from scattering arguments by 't Hooft in ref.[10] and
is implicit in several earlier works]. Given such a density of 
state, these degrees of freedom will possess a relation between mean 
energy and temperature which is appropriate for that of a blackhole.
The excitation and subsequent de-excitation of these degrees of freedom with
 high energy modes of other fields will allow for the other fields to reach
thermal equilibrium at the characteristic temperature. The entire picture
seems to be quite consistent provided the density of state can be 
understood from some more fundamental considerations.

The situation is --- in fact --- reminiscent of the fact that a classical
blackhole is describable by just three parameters, irrespective of the
details of the original system which collapsed to form the it. Similarly,
the quantum blackhole has a universal form of density of states which is
independent of the form of the density of states of the original system
which collapsed to form the blackhole. We will see in the next section that
this feature allows us to construct several model systems with the correct
form of density of states for the blackhole.

The above arguments notwithstanding, one may not be quite happy with 
any physical system which has a density of states which grows so 
fast. In judging such a prejudice, we have to keep the following 
caveats in mind: 
(i) There exists no really 
 not a good argument against  such density 
of states for self gravitating systems which  are, anyway, not describable 
by canonical ensemble. The partition function for such systems cannot
be obtained as a Laplace transform of density of states and has 
to be defined in a generalized manner using some suitable contour 
in complex plane. This is precisely what we have done above.
(ii) The existence of a density of state for a dynamical
system, by itself, does not mean much. Consider for example, the density
of states for an ordinary solid. While formal integrations
over $E$ extending up to infinity will be done in defining variables
like partition function, it is always assumed that high energy physics
is irrelevant if those modes are not excited. So in studying a solid
at room temperatures, we
are not bothered by the form of $g(E)$ for say, $E>1 $ GeV.  
This works well for systems in which mean energy is comparable to the 
temperature. In the case of a stellar mass blackhole, the key difference
is that  the mean energy 
and temperature are widely different and hence it is not clear
up to what energy scales one needs to know the density of states.
 (iii) The above comment is particularly relevant because one 
cannot use thermodynamic description for arbitrarily high energies. 
Eventually the details of the microscopic degrees of freedom, whatever
they are, will be brought into picture. Of course, when this happens, 
even the concept of density of states may break down and we enter
a truly unknown region of quantum gravity.

\section {Modelling the blackhole density of states}

Since $g(E)$ for the blackhole contains the essence of its
behaviour, let us ask which physical systems possess such a density 
of states. While attempting to do this, it is necessary to note
that the density of states obtained above can only be trusted for $E\gg
E_P$. 
In general, the density of 
states can have a form 
\begin{equation}
\rho(\bar E) \approx \exp[4\pi (\bar E/E_P)^2+{\cal O}(\ln(\bar E/E_P) \cdots]\label{eqn:qrhoneeded}
\end{equation}
 where the leading log corrections are unimportant
for $\bar E\gg E_P$. I will now show how several toy field theoretical 
models can be constructed, which have such a density of state. While this 
shows that blackhole entropy by itself cannot provide more information 
regarding the quantum micro-structure, the study of the model systems
will reveal some interesting features. 

The most straight forward attempts to obtain such density of state are
based on combinotrics and using expansions of the form:
\begin{equation}
g(E)=\exp\left({4\pi M^2\over L_P^2}\right)=\exp\left({{\cal A}\over 4L_P^2}
\right)=\sum_{n=0}^\infty{1\over n!}\left({\cal A}\over 4 L_P^2\right)^n
\approx\lim_{N\to\infty}\left({\cal A}\over 4 L_P^2N\right)^N
\end{equation}
where ${\cal A}$ is the area of the event horizon. These expansions
readily suggest interpretations based on combinotrics in which the event horizon is
divided into patches of Planck size. We shall not consider them since it
seems to me that they do not have a direct dynamical content.

To obtain a model for blackhole density of states with some dynamical
content, we can take a clue from the arguments presented in section 3. 
Consider any physical system with a Hamiltonian $H(\Gamma)$
which depends on some suitably defined phase space variables denoted 
symbolically by $\Gamma$. When the system has energy $E$,  let the 
density of states be given by 
\begin{equation} 
g(E) = \int d\Gamma\ \delta_D \left[ E - H(\Gamma)\right]
\propto E^N
\end{equation}
where $N$ is related to the degrees of freedom of the system. We now let  this system collapse in a spherically symmetric manner to form a 
blackhole with the same energy $E$. Once the blackhole is formed,  the density of states should have the characteristic form
$g_{\rm BH}(E) = \exp(4\pi E^2/E_P^2)$ for $E\gg E_P$. If $E$ remains conserved during 
the collapse, then this can only happen if we assume that the formation
of the event horizon somehow changes the Hamiltonian of the system to the 
form 
\begin{equation} H_{\rm mod}^2=A^2\ln\left(1+{H^2\over A^2}\right); \qquad A\propto E_P(N+1)^{1/2}\label{eqn:qpth}
\end{equation}
This is easy to verify: The density of states for $H_{\rm mod}\equiv
f(H)$, where $f^2(x)=A^2\ln[1+(x^2/A^2)]$, is given by
\begin{eqnarray}
 g_{\rm mod}(E) &  & = \int d\Gamma\, \delta_D \left( f[H(\Gamma)] - E\right) = \int d\Gamma \int_0^\infty d\epsilon \, \delta_D \left( H(\Gamma) - \epsilon\right) \, \delta_D\left( f(\epsilon) - E\right)\nonumber\\
& & = \displaystyle{\int_0^\infty d\epsilon \, g(\epsilon) \, \delta_D \left( f(\epsilon) - E\right) = \left[ g( \epsilon) \Big|{df\over d\epsilon}\Big|^{-1}\right]_{\epsilon=\epsilon_r(E)}}
\end{eqnarray}
where $\epsilon_r(E)$ is the root of the equation $f(\epsilon_r)=E$. We have,
\begin{equation}
 \left[ g( \epsilon) \Big|{df\over d\epsilon}\Big|^{-1}\right]_{\epsilon=\epsilon_r} = \left( {E\over A} \exp{E^2\over 2A^2}\right) \left[ A^{N} \left(\exp {E^2\over A^2} - 1 \right)^{N/2} \right] \simeq \exp\left[ {N+1\over 2A^2} E^2 + {\cal O} (\ln E)\right] 
\end{equation}
for $E\gg E_P$. This has the correct form for the blackhole density of
states if we take $A^2=[(N+1)/8\pi]E_P^2$.

If the effect of quantum structure of spacetime is to modify {\it any}
low energy hamiltonian in such a form, then any physical system will have
the density of states needed for blackholes for  $E\gg E_P$. Such a modification will also ensure that all the symmetries of the $H_{conv}$ will be
preserved by $H_{mod}$. Consider, for example, the following question: Some physical system, say, a spherical cloud of dust has a certain hamiltonian
$H_{conv}$ and energy $E$ at $t=0$. It will have certain amount of phase
volume available corresponding to this particular hamiltonian and energy.
When this system collapses to form a blackhole, its energy is conserved but
the density of states have to change to a particular form. One possible way
of modeling this transition is to find a general principle which will modify
the hamiltonian of any system which forms a black hole in the manner suggested
above.
We have no idea, right now, as to how this transformation takes place and 
as such, it should be thought of as a completely ad-hoc postulate.
Right now I will merely use this postulate to construct model systems
with the correct form of density of states for the blackhole. Towards
the end of this section, I will discuss further consequences of this
postulate. I stress that whether such a {\it universal} postulate is valid 
or not is immaterial for the {\it  specific} models which I construct
below.

The simplest {\it dynamical} system which has the required density of states
can be constructed as follows: Consider some system with the Hamiltonian
$H(p)$ which only depends on the magnitude of a D-dimensional momentum
vector ${\bf p}$. The phase volume $\Gamma[E]$ bounded by the energy surface $H(p)=E$,
in the momentum space is given by
\begin{equation}
\Gamma(E)=\int d^Dp\ \Theta(E-H(p))\propto p_{\rm max}^D(E)
\end{equation}
where $p_{max}$ is the momentum corresponding to the energy $E$. For $E\gg E_P$
we want this phase volume to grow as $\exp(4\pi E^2/E_P^2)$. [It does not matter
whether we work with $\Gamma[E]$ or $g(E)$ for 
$E\gg E_P$ since they differ only by sub-dominant logarithmic terms mentioned
in equation (\ref{eqn:qrhoneeded}).]. It follows that $E^2\to (E_P^2 D/8\pi)\ln p_{max}^2$
in this limit. Hence the Hamiltonian we need has the asymptotic form
$H(p)\to (E_P^2 D/8\pi)\ln p^2$ in the same limit. The form of the hamiltonian
for $E\ll E_P$ is not fixed by these considerations but we will expect it to
reduce to some conventional form independent of $L_P$. One simple choice
we will explore is
\begin{equation}
 H^2(p)={E_P^2 D\over 8\pi}\ln\left(1+{8\pi p^2\over DE_P^2}\right)
\label{eqn:qhtent}
\end{equation}
which reduces to that of a massless particle in the low energy limit.
 [The expression in (\ref{eqn:qhtent})
is of the form a relation between a modified Hamiltonian $H_{\rm mod}$
and the conventional low energy Hamiltonian $H_{\rm conv}$ postulated above.
Using $(p^2+m^2)$ in place of $p^2$ in the above expression we can arrange
for the low energy limit to describe a {\it massive} particle; we will not
bother to do this here since we anyway expect $m\ll E_P$.]

Classically, such a system describes particles which move with trajectories
of the form
\begin{equation}
{\bf x}={\bf v}t;\qquad  {\bf v}={\partial H\over \partial{\bf p}}=
{{\bf p}\over E({\bf p})}{1\over [1+(8\pi p^2/DE_P^2)]}\label{eqn:qgrvel}
\end{equation}
For  $E\ll E_P$, this is just a null ray of massless particle; but for
 $E\gg E_P$ we have $v\to [1/(p\ln p)]$ which decreases with increasing momentum. One may interpret this as the dynamics slowing down significantly
at high energies. Quantum mechanically, we get the modified Klein-Gordon
equation for a scalar field to be
\begin{equation}
{\partial^2\phi\over\partial t^2}+A^2\ln\left(1-{\nabla^2\over A^2}\right)\phi=0;
\quad A^2={D\over 8\pi}E_P^2
\end{equation}
The general solution is of the form
\begin{equation}
\phi(t,{\bf x}) = \int {d^Dk\over (2\pi)^D} \, a({\bf k}) \exp (- i\omega_{\bf k} t + i {\bf k\cdot x})
\end{equation}

with the dispersion relation 
\begin{equation} 
\omega^2({\bf k}) = A^2 \ln \left( 1 + {k^2 \over A^2} \right) ; \qquad A^2 = {D\over 8\pi} E_P^2 \label{eqn:qdisp}
\end{equation}
[The $a({\bf k})$ gives the initial amplitude of mode]. Any wave packet
will, of course, disperse under such an evolution. The low-k modes move with
the speed of light while the high-k modes do not propagate at all since their
group velocity [given by equation (\ref{eqn:qgrvel})] grinds to zero at high-k.

The above attempts, of course, are based on ``single particle'' models
and hence are probably too naive. It is however easy to construct
a field theory such that the one-particle excited state of the theory
can have the same density of states as the blackhole. To do this, we only
have to put together a bunch of harmonic oscillators with the dispersion relation given by (\ref{eqn:qdisp}). More formally, let us assume that the transition to continuum spacetime limit of a blackhole can be described
in terms of certain fields  $\phi(t, {\bf x})$  which are  to be constructed
in some suitable manner from the fundamental microscopic variables $q_j$. 
 I take the Lagrangian describing the effective field 
$\phi(t,{\bf x})$ to be 
\begin{equation}
L = {1\over 2} \int d^D{\bf x} \, \dot \phi^2 - {1\over 2} \int d^D{\bf x} d^D{\bf y}\, \phi({\bf x}) F({\bf x} - {\bf y}) \phi({\bf y}) = \int {d^D{\bf k}\over (2\pi)^D} {1\over2} \left[ |\dot Q_{\bf k}|^2 - \omega_{\bf k}^2 |Q_{\bf k}|^2\right]\label{eqn:qlag}
\end{equation}
The Lagrangian is non local in the space coordinates ${\bf x}$ which is taken to be $D-$dimensional; the corresponding Fourier space coordinates are labelled
by ${\bf k}$. The quadratic non locality allows us to describe the system
in terms of free harmonic oscillators with a dispersion relation $\omega({\bf k})$ related to $F({\bf r}) $ by 
\begin{equation}
 \omega^2({\bf k}) = \int d^D{\bf r} \, F({\bf r}) e^{i{\bf k}\cdot{\bf r}}\label{eqn:qomf}
\end{equation}
The energy levels of the system are built out of elementary excitations
with energy $\hbar \omega({\bf k})$ and the density of states corresponding
to the one particle state of the system is:
\begin{equation}
g(E) = \int d^D k \, \delta_D \left[ \omega({\bf k}) - E \right] 
\end{equation}
which is just the Jacobian  $|d^D{\bf k}/d\omega|$. It is 
straightforward to see that if the dispersion relation $\omega({\bf k})$ has
the asymptotic form 
\begin{equation}
\omega^2({\bf k}) \to {E_P^2 D\over 8\pi} \ln k^2 \qquad ({\rm for}\ k^2 \gg E_P^2)\label{eqn:qasydis}
\end{equation}
then the density of states has the required form for that of a blackhole.

In this class of model, the blackhole is treated as the one particle state
of a {\it nonlocal} field theory. The dispersion relation is arranged so as to
give a normal massless scalar field at low energies. It is certainly of interest to explore the
new features of this field theory. 
Since non-locality is the key new feature let us begin by studying the form
of $F({\bf x})$ in real space. We can evaluate it by inverting the
Fourier transform in (\ref{eqn:qomf}) and using (\ref{eqn:qdisp}). The logarithmic singularity
can be regularized by using an integral representation for ln and interchanging the orders of integration. These steps lead to
\begin{eqnarray} 
 F({\bf x})  & & = \displaystyle{\int {d^D{\bf k}\over (2\pi)^D} \, A^2 \ln \left( 1 + {k^2 \over A^2}\right) \, e^{i{\bf k\cdot x}} = - A^2 \int  {d^D{\bf k}\over (2\pi)^D} \int_0^\infty {d\mu \over \mu} e^{-\mu - {\mu\over A^2} k^2 + i {\bf k\cdot x}}\nonumber}\\
& & = \displaystyle{- {A^{2 +D}\over (4\pi)^{D/2}} \int_0^\infty {d\mu\over \mu^{1+D/2}} \, e^{-\mu - {A^2x^2\over 4\mu}} = - {2A^{2+D}\over (2\pi)^{D/2}} \left({1\over A^2x^2}\right)^{D/4} \, K_{D/2} \, \left( Ax\right); \quad A^2 = {D\over 8\pi}\, {1\over L_P^2}}
\label{eqn:qfofx}
\end{eqnarray}
where $K_\nu(z)$ is the MacDonald function. This function vanishes exponentially
for large values of the argument, showing that the effective correlation 
length of
the nonlocal field is of the order of $A^{-1}=(8\pi/D)^{1/2}L_P$. For small
$z$, we have  $K_\nu(z)\to z^{-\nu}$ and the correlation function goes as
a power law $F(x)\propto x^{-D}$. The short distance behaviour of the 
correlation function is universal and depends only on the asymptotic form
of the dispersion relation. In fact, this is the only feature  of $F({\bf x})$ which is needed 
to reproduce the  density of states leading to the correct theory 
of blackhole thermodynamics.  When $L_P \to 0$, the function $F(x)$ is
proportional to the second derivative of Dirac delta function as can be 
seen from the fact that, as $L_P \to 0, \omega^2(k) \to k^2$. In this 
limit, we recover the standard local field theory. 

The expressions are more tractable for the
simplest case of $D=1$. I will now illustrate the above phenomena using the simplest possible choice,
corresponding to $D=1$ and a dispersion relation 
\begin{equation}
\omega^2(k) = {E_P^2\over 8\pi} \ln \left( 1 + {8\pi k^2\over E_P^2}\right) \label{eqn:qdispch}
\end{equation}
The density of 
states corresponding to this dispersion relation is given by 
\begin{equation}
g(\bar E) \cong  \exp\left[ 4\pi {\bar E^2\over E^2_P} + {\cal O} \left( \ln {\bar E\over E_P} \right) \right] \equiv \exp S(\bar E) 
\end{equation}
The corresponding blackhole temperature is 
\begin{equation}
T(\bar E) = \left( {\partial S\over \partial \bar E}\right)^{-1} = {E_P^2 \over 8\pi \bar E} \left[ 1 + {\cal O} \left({ E_P^2\over \bar E^2}\right) \right] \cong {E_P^2\over 8\pi M} 
\end{equation}
for $\bar E = M \gg E_P$.

The function $F( r)$ corresponding to the $\omega^2( k)$ in equation (\ref{eqn:qdispch})  is 
\begin{equation}
F(x) = {E_P^2\over 8\pi}\int^\infty_{-\infty} {dk\over 2\pi} e^{-ikx} \ln \left( 1 + {8\pi k^2\over E_P^2}\right) = -{E_P^3\over 8\pi} \left( {L_P\over |x|}\right) \exp \left( - {|x|\over \sqrt{8\pi} L_P}\right)
\end{equation}
for finite, nonzero, $x$. This can obtained by direct integration, or more
simply, from (\ref{eqn:qfofx}) using  the identity $K_{1/2}(z)=(\pi/2z)^{-1/2}\exp (-z)$.
The functional form of $F(x)$ clearly illustrates the smearing of the 
fields over a region with correlation length $\sqrt{8\pi}\, L_P$. 

The physical content of the above analysis can be viewed as follows.
I start with certain loosely defined dynamical variables $q_j$ describing
the quantum micro-structure of the spacetime. The dynamical theory
describing $q_j$'s must lead, in suitable limit, to a continuum
spacetime with quantum states having mean energies much larger
than $E_P$. Among them are the classical spacetimes
with compact, infinite redshift surfaces like that of a blackhole forming out of a 
stellar collapse. I describe these blackhole spacetimes
in terms of an intermediate effective field theory in $(D+1)$ dimension. The existence of an infinite redshift surface 
allows the elementary excitations of this field with arbitrarily
high energies to occur in such spacetimes. 
Such a theory is nonlocal in space and is based on smearing of fields
over a correlation length of the order of $L_P$. The one particle excitations of this field has the correct density of states to describe the black hole.

The above approach also
suggests that there is nothing mysterious in completely different
microscopic models (like those based on strings or Ashtekar variables)
leading to similar results regarding blackhole entropy. Any theory 
which has the correct density of states can do this; in fact, the 
models I have described are only a very specific subset of several such
toy field theories which can be constructed. The situation is reminiscent
of one's attempt to understand the quantum nature of light
from blackbody radiation. The spectral form of blackbody radiation 
can be derived from the assumption that $E=\hbar \omega$ and is 
quite independent of the details of quantum dynamics of the electromagnetic 
field. Similarly, the blackhole thermodynamics can be explained if 
one treats spacetimes with event horizons as highly excited states of a 
toy, nonlocal field theory whose elementary excitations obey a dispersion 
relation with the asymptotic form given by (\ref{eqn:qasydis}).

 Within the limited point of view of modeling a blackhole,  one need not even identify the $(D+1)$ dimensional space
as a superset of conventional spacetime. The ${\bf x}$ and $t$ could
represent variables in some abstract space and the spacetime structure
could emerge in a more complicated manner in terms of the fields themselves.
However, if we take the postulate of (\ref{eqn:qpth}) seriously, then it is 
logical to think of the $(D+1)$ dimensional space to be connected 
with spacetime in some manner. In that case, the field theoretic
model discussed above has no Lorentz invariance at length scales
comparable to Planck length. This does not bother me in the 
least and, in fact, I consider the breakdown of Lorentz invariance
at Planck scales almost inevitable. There are physical reasons to 
believe that one cannot measure length scales with an accuracy greater than
Planck length which suggests that some of the basic postulates 
of special relativity like the propagation of light signals to identify
spacetime events etc. will get modified at very small length scales. Lorentz
invariance emerges as a symmetry of the continuum spacetime in the models
constructed above and hence all the conventional 
experimental consequences of special relativity, of course, are preserved.
The situation is analogous to the emergence of smooth continuum description 
of a solid from a discrete crystal lattice. The continuum system
can possess translational and rotational invariance for infinitesimal
translations and rotations; but the microscopic crystal lattice will not
respect these symmetries. Similarly, the lattice structure of the quantum spacetime can break the continuum symmetry of Lorentz group. One, of course,
needs to study the modified structure which emerges from the 
above postulate but I am not certain whether this will lead to 
any unique microscopic description. The ideas presented here
also has connections with the principle of path integral duality [14] which 
I have discussed in detail elsewhere. 

Let me now return to the principle of transmutation of Hamiltonian 
encoded in the equation (\ref{eqn:qpth}). Such a postulate can be viewed at different
levels depending on ones philosophical inclination. We know that,
classically, blackhole spacetimes are characterized by very few 
parameters and --- in the context of Schwarzschild blackhole --- energy is the
only relevant quantity. In the quantum mechanical context, the properties of
such a blackhole is encoded in the density of states which has a universal form
independent of the original physical system from which the black hole has
formed. One way of incorporating this universality will be to postulate that
quantum dynamics of blackholes leads to the transmutation of Hamiltonians
as suggested by (\ref{eqn:qpth}). In the  case of spacetimes with a timelike Killing vector $\xi^a$, the Hamiltonian $H$ can be defined as a generally covariant scalar as
\begin{equation}
H = \int T_{ab} \xi^a d \sigma^b
\end{equation}
where $T_{ab}$ is the stress tensor of matter and $d \sigma^b$ is the element of volume on a spacelike surface. According to equation (\ref{eqn:qpth}), this $H$ is modified to
\begin{equation}
H^2_{\rm mod} = A^2 \ln \left(1 + {1 \over A^2} \left[ \int T_{ab} \xi^a d \sigma^b \right]^2 \right)
\end{equation}
In this approach, $H_{\rm mod}$ remains a covariant scalar. I plan to discuss the consequences of this modification elsewhere.

\section {Conclusions}

Since the point of view advocated in this paper has been described `online'
I shall merely summarize the key results in this section. We begin by 
assuming that: 

(i) There exists certain microscopic degrees of 
freedom for the spacetime which manifest themselves only at length scales
comparable to $L_P$. The continuum description of spacetime as a 
solution to Einstein's equation is an approximate one and is similar
to the description of a solid by laws of elasticity. In the 
description of any macroscopic spacetime by some parameters, one
is coarse-graining over a large number of microscopic configurations
of the quantum spacetime. 

(ii) Just as one can provide a thermodynamical description of 
matter without knowing the details of atomic physics, it is possible 
to provide a semiclassical description of spacetime which is reasonably
independent of the microscopic theory. 

Of these two assumptions, the 
first one seems to be generally accepted by most of the workers in 
the field. The second one should be treated as a working hypothesis at 
the moment. Given these two assumptions, one would like to investigate
situations in which some properties of the  microscopic spacetime structure 
will manifest itself. I have given detailed arguments in sections 3 and 4
as to why event horizons can help us in this task. The study of event 
horizons leads to the conclusion that the density of states of 
blackholes must have a particular form in order to provide the
correct thermodynamic description. In fact, any physical system
with such a density of state will be indistinguishable from a blackhole 
as far as thermodynamic interactions are concerned. 
Since a blackhole can be formed from the collapse of any physical 
system, the above result suggests the possible existence of some of new physical principle in the theory of 
quantum blackholes along the following lines: Physical systems characterized by a given
Hamiltonian will have particular energy dependence for the 
density of states. When such a a system collapses to form a blackhole,
conserving the energy, the density of state has to change to a universal
form. This is possible only if  systems which  collape to form 
blackholes are described by an effective Hamiltonian 
which is related to the original hamiltonian in a particular
manner. Using this principle, it is possible to construct several 
model systems which have the correct density of states for the
blackhole. Most important among them are those based on non local
field theories with particle states  having the same density of 
states as a blackhole. Several features of such field theories are 
discussed in section 6.

While it may not be possible to obtain a unique quantum description of 
spacetime from our knowledge of semiclassical blackhole physics, it 
does give three clear pointers. First is the indirect, but essential, role played by the infinite
redshift surface: It is the existence of such a surface which distinguishes
the {\it star} of mass $M$ from a {\it blackhole} of mass $M$. A stellar spacetime will
not be able to populate high energy states of the toy field as required
by the statistical description; in a blackhole spacetime, virtual modes of
arbitrarily high energies near the event horizon will allow this to occur. 
(It may be possible to model such a process by studying the interaction of
this toy field with a more conventional field near the event horizon.)
Second one is the universal transformation of the hamiltonian of
physical systems which collapses to form the blackhole. This transformation leads to a unique asymptotic form for the 
dispersion relation for the elementary excitations of the model field
theories. The third
is the fact that such a dispersion relation almost invariably
leads to smearing of local fields over regions of the order of Planck
length.

I thank Apoorva Patel  for several stimulating discussions.

\end{document}